\documentclass[final,5p,times,twocolumn]{elsarticle}

\usepackage{amsmath,amssymb}
\DeclareMathOperator{\Tr}{Tr}

\journal{Optics Communications}

\begin{document}

\begin{frontmatter}



\title{Quantum teleportation of qudits by means of generalized quasi-Bell states of light}

\author[1,2]{D.~B.~Horoshko\corref{cor1}}
\cortext[cor1]{Corresponding author}\ead{dmitri.horoshko@univ-lille.fr}
\author[1]{G.~Patera}
\author[1]{M.~I.~Kolobov}

\address[1]{Univ. Lille, CNRS, UMR 8523 - PhLAM - Physique des Lasers Atomes et Mol\'ecules, F-59000 Lille, France}
\address[2]{B.~I.~Stepanov Institute of Physics, NASB, Nezavisimosti Ave.~68, Minsk 220072 Belarus}

\begin{abstract}
Quantum superpositions of coherent states are produced both in microwave and optical domains, and are considered realizations of the famous ``Schr\"odinger cat'' state. The recent progress shows an increase in the number of components and the number of modes involved. Our work creates a theoretical framework for treatment of multicomponent two-mode Schr\"odinger cat states. We consider a class of single-mode states, which are superpositions of $N$ coherent states lying on a circle in the phase space. In this class we consider an orthonormal basis created by rotationally-invariant circular states (RICS). A two-mode extension of this basis is created by splitting a single-mode RICS on a balanced beam-splitter. We show that these states are generalizations of Bell states of two qubits to the case of $N$-level systems encoded into superpositions of coherent states on the circle, and we propose for them the name of generalized quasi-Bell states. We show that using a state of this class as a shared resource, one can teleport a superposition of coherent states on the circle (a qudit). Differently from some other existing protocols of quantum teleportation, the proposed protocol provides the unit fidelity for all input states of the qudit. We calculate the probability of success for this type of teleportation and show that it approaches unity for the average number of photons in one component above $N^2$. Thus, the teleportation protocol can be made unit-fidelity and deterministic at finite resources.
\end{abstract}


\end{frontmatter}


\section{Introduction}\label{sec:intro}  

Superpositions of two coherent states of an optical mode for several decades attract attention of the scientific community as optical realizations of the famous ``Schr\"odinger cat'' state of a quantum system \cite{Dodonov74,Buzek92,Brune96,Ourjoumtsev06,Nielsen06,Sychev17,Yurke86,Kirchmair13,Vlastakis13}. There are two classes of these states with distinct properties and generation methods: (i) ``even'' and ``odd''  coherent states \cite{Dodonov74,Buzek92,Brune96,Ourjoumtsev06,Nielsen06,Sychev17}, and (ii) the Yurke-Stoler coherent state \cite{Yurke86,Kirchmair13,Vlastakis13}. The states of the first class contain either even or odd number of photons and are generated for high-Q microwave cavity field by interaction with non-resonant Rydberg atoms \cite{Brune96} or for traveling-wave optical field by photon subtraction from a squeezed state \cite{Ourjoumtsev06,Nielsen06,Sychev17}. The states of the second class have a Poissonian distribution of photons and have been recently generated for microwave cavity field coupled to a superconducting qubit by nonlinear Kerr effect \cite{Kirchmair13,Vlastakis13}. The states of both classes have been extensively studied as models of decoherence \cite{Buzek92,Brune96,Horoshko98}, sources of quantum instabilities \cite{Kilin96,Horoshko00} and resources for quantum computation \cite{Jeong02,Ralph03,Mirrahimi14,Albert16}.

Splitting these superpositions in two modes, one obtains entangled coherent states \cite{Sanders92,Hirota01,vanEnk01,Jeong01,Joo11,Reut17,Wang16}. The states of the first class create in this way quasi-Bell states \cite{Hirota01}, having the same structure as usual Bell states of two qubits, but with two non-orthogonal coherent states as basis for each mode. These states can be applied to quantum metrology \cite{Joo11}, quantum tomography \cite{Reut17} and probabilistic quantum teleportation \cite{vanEnk01,Jeong01}, which is a key element of coherent state quantum computation. While traveling-wave superpositions can be rather easily split by a beam-splitter, cavity fields require a more sophisticated technique, realized only recently for two coupled microwave cavities \cite{Wang16}.

Increasing the number $N$ of coherent components in a single-mode superposition, one obtains ``multiple component Schr\"odinger cats'' \cite{HarocheRaimond06}. The most attention has been attracted to coherent states placed equidistantly on a circle \cite{HarocheRaimond06,Janszky93,Domokos94}, though other geometries are also possible \cite{Kilin95}. Multicomponent extensions of the second class \cite{Tanas91} are known as ``Kerr states'' and can be generated by the same Kerr effect as their two-component variants, as has been  demonstrated recently in the microwave domain \cite{Kirchmair13,Vlastakis13}. Production of multicomponent states of the first class has been proposed for cavity field \cite{Domokos94}, but not yet reported. These states are highly important for studying  decoherence \cite{Zurek01}, and for application in quantum computation with qudits, quantum systems with the number of levels higher than two \cite{Kim15,Li17}.

Splitting multiple component Schr\"odinger cat states in two modes, one naturally arrives at entangled coherent states of high dimension \cite{vanEnk03,Kilin11}. The states of the first class create in this way the states which are generalizations of quasi-Bell states, and their structure has been recently analyzed in detail \cite{Horoshko16}. These states have been proposed for enhancing sensitivity in quantum metrology \cite{Lee15}. The states of the second class, entangled Kerr states, can be used for quantum teleportation of high dimensional systems \cite{vanEnk03}.

In this work we further develop the approach of Ref.~\cite{Horoshko16} and introduce explicitly entangled states of two optical modes, which we call ``generalized quasi-Bell states'' (GQBS). They are ``quasi-Bell'' because their basis vectors are not exactly orthogonal, the term having been coined in Ref.~\cite{Hirota01} for the $N=2$ case. At higher $N$ they generalise the quasi-Bell states to higher dimensions like generalized Bell states \cite{Cerf00,Horoshko07} do for the Bell states. GQBS create a natural basis for quantum information processing with coherent states of light. In particular, we show that the protocol of probabilistic quantum teleportation works for these states much better than for the entangled Kerr states, for which it was originally suggested \cite{vanEnk03}. We show that the GQBS-based protocol of quantum teleportation can provide, under certain conditions, an almost unit probability and almost unit fidelity at finite resources.

\section{Generalized quasi-Bell states}
\begin{figure}[h]
\begin{center}
\includegraphics[width=0.7\columnwidth]{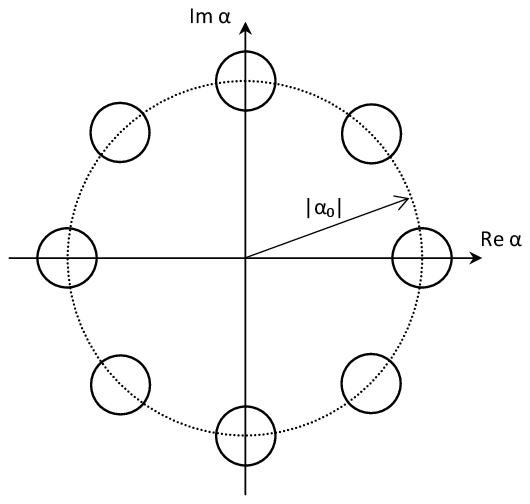}
\caption{\label{Fig1} Coherent states on the circle of radius $|\alpha_0|$. Each coherent state is represented by a circle of radius $\frac12$, corresponding to the $\sigma$-area of its Wigner function.}
\end{center}
\end{figure}

We consider one mode of the electromagnetic field, for which we fix a set of $N$ coherent states $\{|\alpha_m\rangle, m=0,1,...,N-1\}$, placed equidistantly on the circle of radius $|\alpha_0|$, i.e., $\alpha_m=\alpha_0 e^{-i2\pi m/N}$, where $\alpha_0$ is an arbitrary non-zero complex number  (see Fig.\ref{Fig1}).

We are interested in coherent superpositions of these states having equal weights and linear periodic relative phase:
\begin{equation}\label{RICS}
|\mathrm{c}_q\rangle =  \frac1{N\sqrt{\tilde g(q)}}\sum_{m=0}^{N-1}e^{i2\pi mq/N}|\alpha_0 e^{-i2\pi m/N}\rangle,
\end{equation}
where $\tilde g(q)$ is the discrete Fourier transform,
\begin{equation}\label{tildeg}
\tilde g(k)=\frac1{N}\sum_{m=0}^{N-1}g(m)e^{-i2\pi km/N}=e^{-|\alpha_0|^2}\sum_{l=0}^\infty \frac{|\alpha_0|^{2(k+lN)}}{(k+lN)!},
\end{equation}
of the first column of the Gram matrix $G_{mn}=\langle \alpha_m|\alpha_n\rangle = g(m-n)$, defined as
\begin{equation}\label{g}
g(m)=\exp\left\{|\alpha_0|^2\left(e^{i2\pi m/N}-1\right)\right\}.
\end{equation}
Each state of the set  $\{|\alpha_m\rangle, m=0,1,...,N-1\}$ is produced from the previous one in this set by a rotation in phase space  $|\alpha_m\rangle=U_N|\alpha_{m-1}\rangle$, described by the unitary operator
\begin{equation}\label{U}
U_N = e^{-i2\pi a^\dagger a/N},
\end{equation}
where $a$ is the photon annihilation operator of the optical mode. The eigenvalues of $U_N$ are given by $e^{-i2\pi q/N}$. There are only $N$ different eigenvalues, for which we let $0\le q \le N-1$. The corresponding eigenstates are given by Eq.(\ref{RICS}):
\begin{equation}\label{eigen}
U_N|\mathrm{c}_q\rangle = e^{-i2\pi q/N}|\mathrm{c}_q\rangle.
\end{equation}
The last property allows us to call them rotationally-invariant circular sates (RICS). They satisfy the orthonormality condition $\langle \mathrm{c}_q|\mathrm{c}_r\rangle = \delta_{qr}$. Using the decomposition of a coherent state in the Fock basis, we rewrite the RICS, Eq.(\ref{RICS}), in the form \cite{Janszky93}
\begin{equation}\label{cq}
|\mathrm{c}_q\rangle =  \frac{e^{-|\alpha_0|^2/2}}{N\sqrt{\tilde g(q)}}\sum_{l=0}^{\infty}\frac{\alpha_0^{q+lN}}{\sqrt{(q+lN)!}}|q+lN\rangle,
\end{equation}
where $|q+lN\rangle$ is a Fock state with $q+lN$ photons. Eq.(\ref{cq}) shows that a RICS is a sum of Fock states with the number $m$ of photons such, that $(m \mod N)=q$. In the case of $N=2$ this property is reduced to a fixed parity of the photon number, peculiar to ``even'' and ``odd'' coherent states. In general, RICS are highly nonclassical, their distance from the set of classical states \cite{Bievre19} is lower-bounded by $\sqrt{1+2\langle n\rangle}-1$, where $\langle n\rangle = \langle\mathrm{c}_q|a^\dagger a|\mathrm{c}_q\rangle$ is the mean photon number. Thus, with growing number of photons these states become harder to create and control. However, as we show, even for low values of the mean photon number they can be used for some protocols of quantum information processing.

The states, Eq.~(\ref{cq}) have been recently considered in the case where all coherent states are almost orthogonal \cite{Kim15}, which implies sufficiently high $|\alpha_0|$ and sufficiently low $N$. In this case they are referred to as ``pseudo-number'' states, while the coherent states $|\alpha_m\rangle$ as the corresponding ``pseudo-phase'' states. They create two complementary bases for an $N$-level system (qudit), encoded into a subspace of optical mode's Hilbert space. Note, that the second basis is non-orthogonal for low values of $\alpha_0$.

Now we consider the RICS $|\mathrm{c}_q\rangle$ with parameters $\{N,\sqrt{2}\alpha_0\}$ at one input of a 50:50 beam-splitter with the vacuum at the other one. The state of the two output modes $A$ and $B$ of the beam-splitter is
\begin{eqnarray}\label{out0}
|\Phi_{q0}\rangle_{AB} &=& \frac{1}{N\sqrt{\tilde g_1(q)}}\sum_{m=0}^{N-1}e^{i2\pi qm/N}|\alpha_m\rangle_A|\alpha_m\rangle_B \\\nonumber
&=& \sum_{k=0}^{N-1}\sqrt{\lambda_k(q)}|\mathrm{c}_k\rangle_A|\mathrm{c}_{q-k}\rangle_B,
\end{eqnarray}
where the second part represents a Schmidt decomposition with the Schmidt coefficients \cite{Horoshko16}
\begin{equation}\label{lambda}
\lambda_k(q) = \frac{\tilde g(k)\tilde g(q-k)}{\tilde g_1(q)},
\end{equation}
and
\begin{equation}\label{normpsi}
\tilde g_1(q) = \sum_{k=0}^{N-1}\tilde g(k)\tilde g(q-k) = \frac1N\sum_{m=0}^{N-1}g^2(m)e^{-i2\pi qm/N},
\end{equation}
from where the summation of the Schmidt coefficients to unity follows. Here and below all indexes and integer arguments are taken modulo $N$.

Now we apply the power $p$ of the rotation operator $U_N^p$ with $0\le p\le N-1$ to the mode $B$ of $|\Phi_{q0}\rangle_{AB}$ and obtain the state
\begin{eqnarray}\label{GQBS}
|\Phi_{qp}\rangle_{AB} &=& \frac{1}{N\sqrt{\tilde g_1(q)}}\sum_{m=0}^{N-1}e^{i2\pi qm/N}|\alpha_m\rangle_A|\alpha_{m+p}\rangle_B \\\nonumber
&=& \sum_{k=0}^{N-1}\sqrt{\lambda_k(q)} e^{-i2\pi(q-k)p/N}|\mathrm{c}_k\rangle_A|\mathrm{c}_{q-k}\rangle_B,
\end{eqnarray}
which is, on the one hand, an extension of the quasi-Bell states \cite{Hirota01} to higher dimensions, and on the other hand, an extension of generalized Bell state \cite{Cerf00,Horoshko07} to non-orthogonal basis, and which we call GQBS. Two forms of this state, presented in Eq.~(\ref{GQBS}), correspond to writing it in the non-orthogonal coherent basis with equal weights, or in the orthogonal RICS basis with non-equal weights.

\begin{figure}[h]
\begin{center}
\includegraphics[width=0.49\columnwidth]{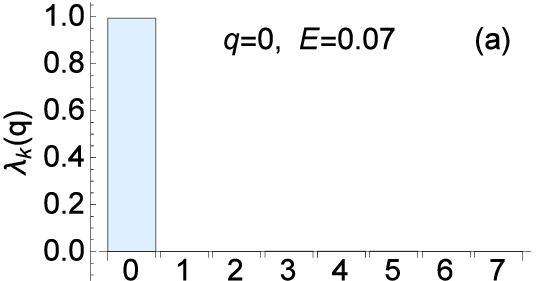}
\includegraphics[width=0.49\columnwidth]{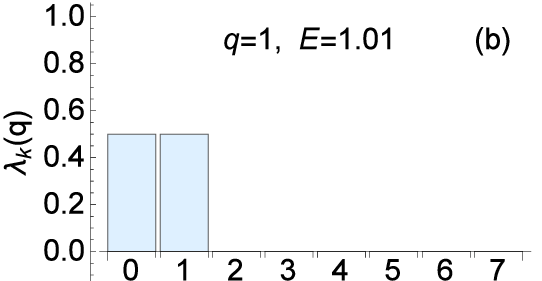}
\includegraphics[width=0.49\columnwidth]{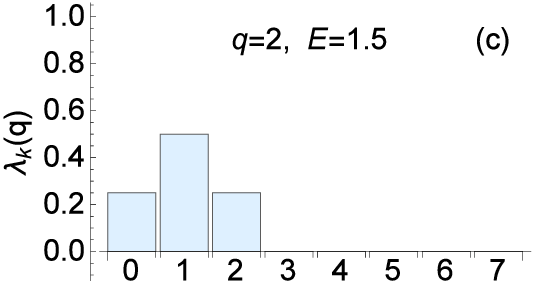}
\includegraphics[width=0.49\columnwidth]{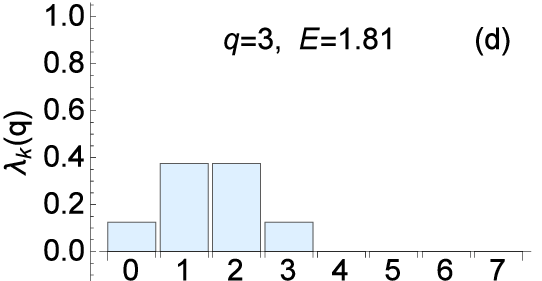}
\includegraphics[width=0.49\columnwidth]{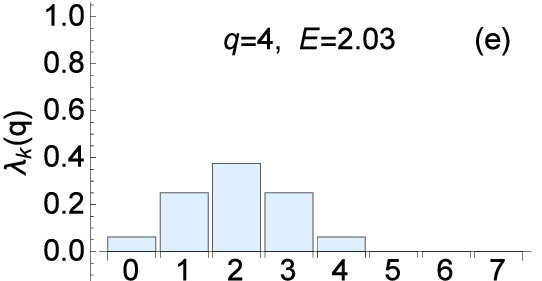}
\includegraphics[width=0.49\columnwidth]{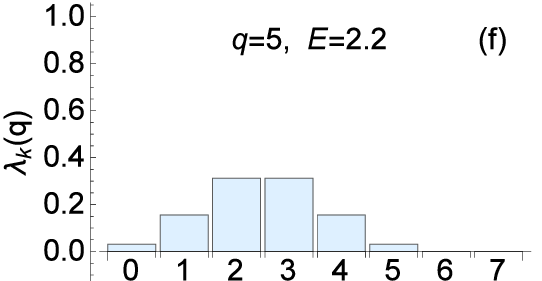}
\includegraphics[width=0.49\columnwidth]{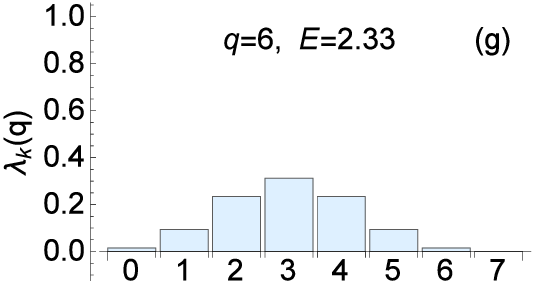}
\includegraphics[width=0.49\columnwidth]{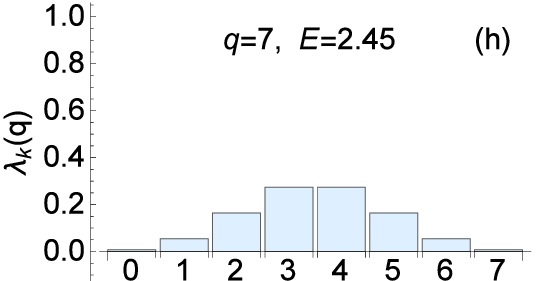}
\caption{\label{Fig2} Schmidt coefficients $\lambda_k(q)$ of a GQBS with $N=8$ and $|\alpha_0|=1$. The entanglement $E$ characterizes the spread of the Schmidt coefficients. Maximal entanglement corresponds to the maximal spread of Schmidt coefficients at $q=7$. For $q=0$ the state is close to the two-mode vacuum with very low entanglement.}
\end{center}
\end{figure}

Averaging out one mode one obtains the reduced density operator of the other one:
\begin{equation}\label{rhoA}
\rho_A(q) = \sum_{k=0}^{N-1}\lambda_k(q)|\mathrm{c}_k\rangle\langle\mathrm{c}_{k}|,
\end{equation}
which tends to the completely undetermined qudit state $\mathbb{I}/N$ in the limit of high $\alpha_0$. In this limit GQBS is maximally entangled, and its entanglement, defined as $E=-\Tr\{\rho_A\log_2\rho_A\}$, is $E=\log_2 N$. For lower values of $\alpha_0$ the reduced state, Eq.~(\ref{rhoA}), is not the completely undetermined one and the GQBS is non-maximally entangled. Entanglement of the state $|\Phi_{qp}\rangle_{AB}$ is given in this case by the Shannon entropy of the Schmidt coefficients $\lambda_k(q)$
\begin{equation}\label{E}
E(q) = -\sum_{k=0}^{N-1}\lambda_k(q)\log_2\lambda_k(q),
\end{equation}
and is a function of $q$ (but not $p$). Schmidt coefficients $\lambda_k(q)$ are shown in Fig.~\ref{Fig2} for the case where each coherent state has on average just 1 photon.

\begin{figure}[h!]
\begin{center}
\includegraphics[width=0.49\columnwidth]{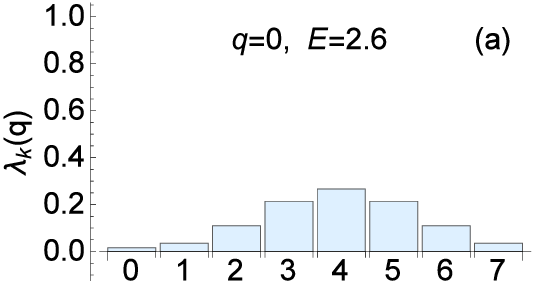}
\includegraphics[width=0.49\columnwidth]{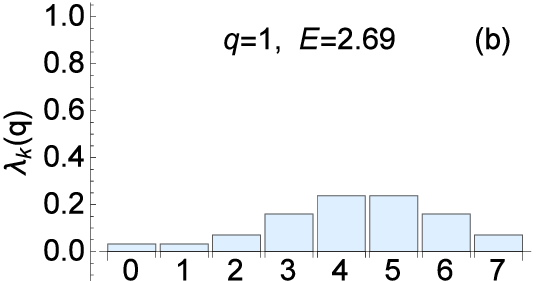}
\includegraphics[width=0.49\columnwidth]{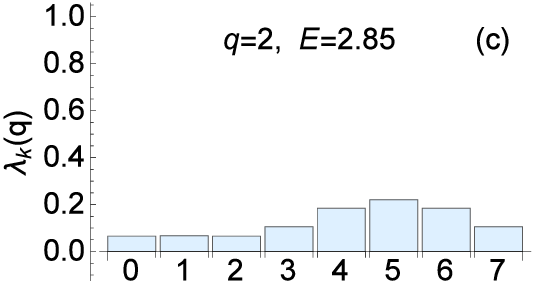}
\includegraphics[width=0.49\columnwidth]{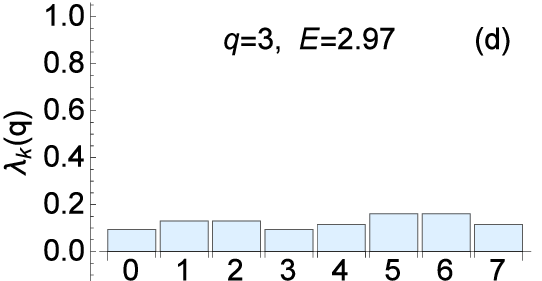}
\includegraphics[width=0.49\columnwidth]{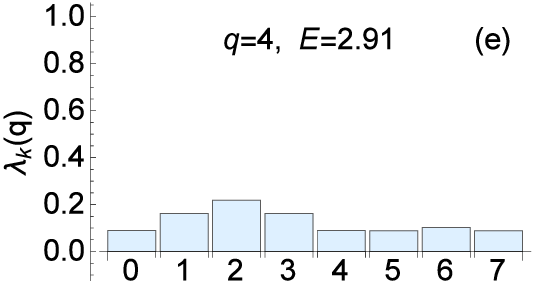}
\includegraphics[width=0.49\columnwidth]{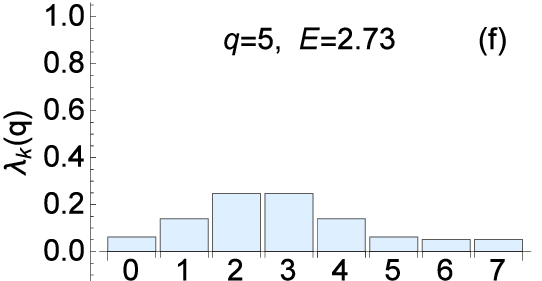}
\includegraphics[width=0.49\columnwidth]{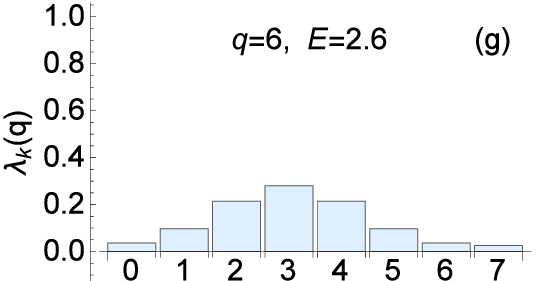}
\includegraphics[width=0.49\columnwidth]{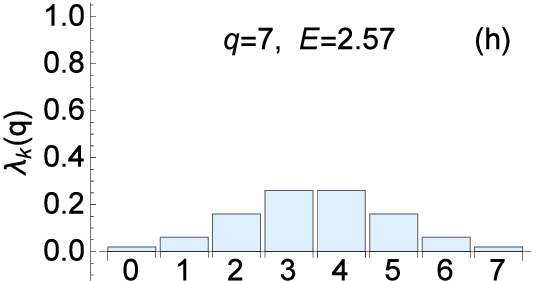}
\caption{\label{Fig3} Schmidt coefficients $\lambda_k(q)$ of a GQBS with $N=8$ and $|\alpha_0|=2$. In this case the maximal spread corresponds to $q=3$, the corresponding state being almost maximally entangled.}
\end{center}
\end{figure}

The spread of coefficients in Fig.~\ref{Fig2} increases with $q$, reaching its maximum at $q=N-1$. Since the entanglement is given by the Shannon entropy of the Schmidt coefficients, it is minimal at $q=0$ and maximal at $q=N-1$. The same coefficients are shown in Fig.~\ref{Fig3} for the case where each coherent state has on average 4 photons. The dependence of entanglement on $q$ is more complicated in this case. The maximum is reached for $q=3$ and the corresponding state $|\Phi_{3p}\rangle$ for any $p$ is very close to a maximally entangled state, which is reflected in its entanglement 2.97, very close to $\log_2N=3$.

\section{Teleportation of circular states of light}

Quantum teleportation of optical states is a key quantum technique using an entangled state of two optical field as a resource \cite{Furusawa98,Horoshko00b}. Entangled two-mode states, GQBS, can be used for teleporting a superposition of states $|\alpha_m\rangle$ with arbitrary coefficients, known as ``circular state'' \cite{Horoshko16} and representing an optical qudit, by the protocol proposed by van Enk \cite{vanEnk03}. This protocol was initially devised for an entangled Kerr state as a resource, but, as we will see, it works even better for a GQBS. Since a GQBS is in general non-maximally-entangled, the teleportation has a non-unit probability of success. Below we reproduce the description of the teleportation protocol from Ref.~\cite{vanEnk03} with a replacement of the nonlocal state.

\textit{
We suppose two parties, Alice and Bob, share a GQBS $|\Phi_{qp}\rangle_{AB}$, given by Eq.~(\ref{GQBS}), with even number of components $N=2L$. Alice possesses in mode $C$ a superposition of coherent states $|\alpha_l\rangle$ with arbitrary coefficients $Q_l$:
\begin{equation}\label{psi}
|\psi\rangle_{C} = \sum_{l=0}^{N-1}Q_l|\alpha_l\rangle_C,
\end{equation}
that she wishes to teleport to Bob.  Alice first uses beam splitters to make $L=N/2$ ``diluted'' copies of both the state to be teleported (ending up in modes $C_k$ for $k=0,...,L-1$) and of her half of the entangled state (ending up in modes $A_k$ for $k=0,...,L-1$) by the process
\begin{equation}\label{dilution}
|\alpha_m\rangle|0\rangle^{\otimes L-1} \to |\alpha_m/\sqrt{L}\rangle^{\otimes L}.
\end{equation}
Then she applies the phase shift operator $U_N^k$ to the mode $A_k$ and, in order to perform her
Bell measurement, subsequently combines the modes $C_k$ and $A_k$ on $L$ 50:50 beam splitters. If we call the output modes $G_k$ and $H_k$ for $k=0,...,L-1$, the resulting state is
\begin{eqnarray}\label{tele}
&&\frac{1}{N\sqrt{\tilde g_1(q)}}\sum_{m=0}^{N-1}\sum_{l=0}^{N-1}e^{i2\pi qm/N}Q_l
|\alpha_{m+p}\rangle_B\\\nonumber
&&\bigotimes_{k=0}^{L-1}|(\alpha_l-\alpha_{m+k})/\sqrt{2L}\rangle_{G_k} \,|(\alpha_l+\alpha_{m+k})/\sqrt{2L}\rangle_{H_k}\,.
\end{eqnarray}
Alice now performs photon-number measurements on all $2L=N$ output modes. She cannot find a nonzero number in every mode. But suppose she finds nonzero numbers of photons in all but one mode, say, mode $H_k$. Then the only terms that survive the sums over $m$ and $l$ in Eq.~(\ref{tele}) are those for which $\alpha_l+\alpha_{m+k} = 0$, that is $e^{-i2\pi l/N}=-e^{-i2\pi(m+k)/N}$ or $m+k-l={L}\mod{N}$. The state at Bob's side reduces to
\begin{equation}\label{tele2}
|\psi'\rangle_{B} = \sum_{l=0}^{N-1}e^{i2\pi q(L+l-k)/N}Q_l|\alpha_{L+l-k+p}\rangle_B.
\end{equation}
Alice communicates to Bob which mode contained no photons, and Bob then applies the appropriate unitary transformation. Here, with $H_k$ being the empty mode, he applies $U_N^{k-p-L}$ to his state to obtain
\begin{equation}\label{tele3}
|\psi''\rangle_{B} = e^{i2\pi q(L-k)/N}\sum_{l=0}^{N-1}e^{i2\pi ql/N}Q_l|\alpha_{l}\rangle_B.
\end{equation}
}

Since the states $|\alpha_{l}\rangle$ are non-orthogonal, the phase factor under the sum cannot in the general case be removed by a unitary transformation. It means that for exact teleportation we need to choose $q=0$ for the entangled state used as resource, which makes Bob's state identical to Alice's, $|\psi''\rangle=|\psi\rangle$. We see also, that the value of $p$ does not affect the process of teleporation and is compensated at the final stage by the rotation operator, so that we can put $p=0$ from the beginning, and consider the state $|\Phi_{00}\rangle_{AB}$ as the optimal resource for quantum teleportation. The analysis of the previous section (see also Ref.~\cite{Horoshko16}) shows that the case of $q=0$ does not always correspond to maximal entanglement for given $N$ and $\alpha_0$, but employment of other GQBS for quantum teleportation is not possible if one requires an exact reproduction of the initial state. Occurrence of zero-photon measurement outcome in more than one mode leads to protocol failure, thus the described protocol is probabilistic.

The possibility of realizing an exact, although probabilistic, teleportation of an arbitrary circular state on the basis of GQBS is a serious advantage compared to the original teleportation scheme \cite{vanEnk03}, where the final state contains phase factors quadratic in $l$, which cannot be corrected by any unitary transformation.

\section{Probability of success}

Teleportation described in the previous section is probabilistic, it succeeds if a zero number of photons is found in one of the measured modes only. If zero photons is found in two or more modes, the protocol fails. The probability of success depends in general on the input state. We show first how this probability can be obtained for a coherent state at the input, and then pass to the general case.

We start with finding the probability of success for the case where the state to be teleported is one of the states $|\alpha_m\rangle$. Thanks to the rotational symmetry, we can choose this state as $|\psi\rangle_C=|\alpha_0\rangle_C$ without loss of generality. The shared entangled state which we consider is $|\Phi_{00}\rangle_{AB}$, the one providing an exact teleportation. For such a choice the multimode state, Eq.~(\ref{tele}), takes the form
\begin{equation}\label{tele2a}
|\Psi\rangle = \frac{1}{N\sqrt{\tilde g_1(0)}}\sum_{m=0}^{N-1}
|\alpha_{m}\rangle_B \bigotimes_{k=0}^{L-1}\left|\frac{\alpha_0-\alpha_{m+k}}{\sqrt{N}}\right\rangle_{G_k} \,\left|\frac{\alpha_0+\alpha_{m+k}}{\sqrt{N}}\right\rangle_{H_k}\,.
\end{equation}

The probability of obtaining zero photons in the mode $H_k$ and non-zero number of photons in the other $N-1$ measured modes is given by the average $\langle\Psi|\Gamma_{H_k}|\Psi\rangle$ of the projector
\begin{equation}\label{Gamma}
\Gamma_{H_k} = \bar\Pi_{G_0}\bar\Pi_{H_0}...\bar\Pi_{G_k}\Pi_{H_k}...\bar\Pi_{G_{L-1}}\bar\Pi_{H_{L-1}},
\end{equation}
where $\Pi=|0\rangle\langle0|$ is the projector on the vacuum, and $\bar\Pi=\mathbb{I}-|0\rangle\langle0|$ is the projector on the non-vacuum state of the mode, whose label is indicated by the lower index. If we denote the summand in the right hand side of Eq.~(\ref{tele2a}) by $|\Psi_m\rangle$, then it is easy to see that only the state $|\Psi_{L-k}\rangle$ makes a non-zero input to $\langle\Psi|\Gamma_{H_k}|\Psi\rangle$. Indeed, each other state has vacuum in a mode different from $H_k$ and gives zero when averaged with the projector $\bar\Pi$ of this mode. Thanks to the symmetry of all $N$ modes the total success probability is $N$ times the probability of the successful outcome in one mode. Finally we find
\begin{eqnarray}\label{P}
P_\mathrm{success} &=& N\langle\Psi_{L-k}|\Gamma_{H_k}|\Psi_{L-k}\rangle\\\nonumber
&=& \frac{1}{N\tilde g_1(0)}\prod_{l=1}^{N-1}\left(1-e^{-|\alpha_0-\alpha_l|^2/N}\right).
\end{eqnarray}

\begin{figure}[h!]
\begin{center}
\includegraphics[width=\columnwidth]{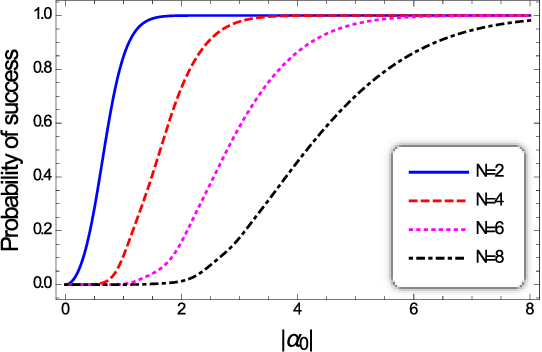}
\caption{\label{fig:P} Probability of success for teleportation of a coherent state as function of the coherent amplitude. For a sufficiently large amplitude $|\alpha_0|$ the coherent states $|\alpha_m\rangle$ become almost orthogonal and the teleportation becomes almost deterministic. However, with the growing number of components $N$, the critical amplitude where the probability approaches 1 grows approximately as $|\alpha_0|\approx N$.}
\end{center}
\end{figure}

This probability is shown in Fig.~\ref{fig:P} for different values of $N$. We see that a practical value 0.2 of probability can be reached for the case of 4 components for $|\alpha_0|\approx1.15$, available in the optical domain \cite{Nielsen06,Sychev17}, and for the case of 8 components for $|\alpha_0|\approx3.1$, available in the microwave domain \cite{Vlastakis13}.

In a similar way we can find the success probability of the original protocol of Ref.~\cite{vanEnk03}. Expression for this probability coincides with Eq.~(\ref{P}) if we omit the normalisation factor before the product. However, since at $|\alpha_0|>1$ we have $N\tilde g_1(0)\approx1$, which means that for practical values of the amplitude the success probability is very close to that of the protocol based on GQBS.

Now we pass to the general case of arbitrary circular state at the input of the teleportation protocol. The state of all modes before the measurement is given by Eq.~(\ref{tele}) with $q=p=0$, which can be rewritten as
\begin{equation}\label{tele3a}
|\Psi\rangle = \sum_{k=0}^{L-1}\left(|\Psi^G_k\rangle+|\Psi^H_k\rangle\right),
\end{equation}
where
\begin{eqnarray}\nonumber
|\Psi^G_k\rangle &=& \frac{1}{N\sqrt{\tilde g_1(0)}}\sum_{l=0}^{N-1}Q_l
\big|\alpha_{l-k}\big\rangle_B \big|0\big\rangle_{G_k}\big|2\alpha_l/\sqrt{N}\big\rangle_{H_k}\\\label{PsiGk}
&&\bigotimes_{j\,\ne\,k}\left|\frac{\alpha_l-\alpha_{l-k+j}}{\sqrt{N}}\right\rangle_{G_j} \,\left|\frac{\alpha_l+\alpha_{l-k+j}}{\sqrt{N}}\right\rangle_{H_j}\,.
\end{eqnarray}
is the state with the vacuum in the mode $G_k$, and
\begin{eqnarray}\nonumber
|\Psi^H_k\rangle &=& \frac{1}{N\sqrt{\tilde g_1(0)}}\sum_{l=0}^{N-1}Q_l
\big|\alpha_{l-k+L}\big\rangle_B \big|2\alpha_l/\sqrt{N}\big\rangle_{G_k}\big|0\big\rangle_{H_k}\\\label{PsiHk}
&&\bigotimes_{j\,\ne\,k}\left|\frac{\alpha_l-\alpha_{l-k+j+L}}{\sqrt{N}}\right\rangle_{G_j} \,\left|\frac{\alpha_l+\alpha_{l-k+j+L}}{\sqrt{N}}\right\rangle_{H_j}\,.
\end{eqnarray}
is the state with the vacuum in the mode $H_k$. The probability of obtaining zero photons in the mode $H_k$ and non-zero number of photons in the other $N-1$ measured modes is given by the average $\langle\Psi|\Gamma_{H_k}|\Psi\rangle$ of the projector, defined by Eq.~(\ref{Gamma}). It is easy to see, that $\langle\Psi|\Gamma_{H_k}|\Psi\rangle = \langle\Psi^H_k|\Gamma_{H_k}|\Psi^H_k\rangle$, since all other states have the vacuum state in one of the modes other than $H_k$, and give zero when averaged with the projector $\bar\Pi$ of this mode. Similarly, we find that the probability of obtaining zero photons in the mode $G_k$ and non-zero number of photons in the other $N-1$ measured modes is given by the average $\langle\Psi^G_k|\Gamma_{G_k}|\Psi^G_k\rangle$ of the projector $\Gamma_{G_k}$, defined by Eq.~(\ref{Gamma}) with exchanged modes $G_k$ and $H_k$.

The total success probability is
\begin{equation}\label{Pgeneral}
P_\mathrm{success} = \sum_{k=0}^{L-1}\left(\langle\Psi^G_k|\Gamma_{G_k}|\Psi^G_k\rangle +\langle\Psi^H_k|\Gamma_{H_k}|\Psi^H_k\rangle\right)
\end{equation}
and can be calculated numerically from Eqs.~(\ref{PsiGk}) and (\ref{PsiHk}) for any input state, defined by the choice of coefficients $Q_l$.

\begin{figure}[h!]
\begin{center}
\includegraphics[width=\columnwidth]{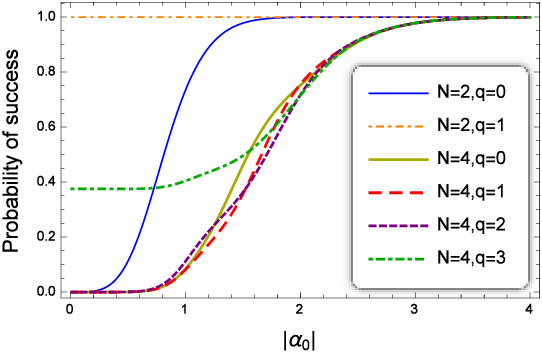}
\caption{\label{fig:Prics} Probability of success for teleportation of a RICS as function of the coherent amplitude. As in the case of coherent state, Fig.~\ref{fig:P}, the teleportation becomes deterministic at $|\alpha_0|\approx N$. At low amplitude the success probability does not tend to zero for $q=N-1$. Moreover, for the RICS with two components and $q=1$, i.e. for the odd coherent state, the teleportation is deterministic at any value of the amplitude (thin orange dot-dashed line).}
\end{center}
\end{figure}

We calculate the success probability for a RICS $|\mathrm{c}_q\rangle$, defined by Eq.~(\ref{RICS}), as an input state. For this state $Q_l=N^{-1}\tilde g(q)^{-1/2}e^{i2\pi ql/N}$. The success probability is shown in Fig.~\ref{fig:Prics} for the cases $N=2$ and $N=4$.

We see that the success probability for the teleportation of RICS tends to unity for $|\alpha_0|$ approaching $N$, as it was for a coherent state. However, its behaviour at $|\alpha_0|\to0$ is not always the same. The success probability tends to zero in the limit of low $\alpha_0$ for all states except for $q=N-1$, where the probability is non-zero at whatever low amplitude. This peculiarity can be understood from Eq.~(\ref{cq}), showing that a RICS  $|\mathrm{c}_q\rangle$ has $q$ photons at minimum, and exactly $q$ photons at $\alpha_0$ approaching zero. Thus, before the measurement the $N$ modes $\{G_k,H_k,k=0,...,L-1\}$ have at least $q$ photons in total. One of these modes is always in the vacuum state due to destructive interference, therefore at least $q$ photons are distributed in $N-1$ modes. When $q<N-1$, the probability to find at least one photon in each of $N-1$ modes tends to zero with the amplitude. But when $q=N-1$ this is not the case, since $N-1$ photons can be found in $N-1$ modes, one photon per mode, and $P_\mathrm{success}$ at $|\alpha_0|\to0$ gives the probability of this event. In particular, for $N=2$, $q=1$ we have at least one photon in one mode, which is necessarily detected (we imply the unit detection efficiency). That is why the teleportation of this particular RICS is deterministic at any amplitude, as shown in Fig.~\ref{fig:Prics}.

To conclude this section, let us compare the teleportation protocol based on GQBS to other continuous variables teleportation protocols \cite{vanEnk03,Furusawa98,Horoshko00b,Andersen13,Kogias14,Marshall14}. The protocol based on the two-mode squeezed state as a shared resource \cite{Furusawa98} is deterministic, but provides a non-unit fidelity. Increase of fidelity to unity requires nonphysical amounts of squeezing and correspondingly energy for the resource state. The same problem takes place for the protocol based on quantum nondemolition interaction \cite{Horoshko00b}, which is also deterministic, but where the unit fidelity can be reached only at an infinitely strong coupling of two optical modes. The protocol of van Enk \cite{vanEnk03}, based on entangled Kerr states, suffers from the same problem: it cannot provide a unit fidelity for all input states, as discussed above. On the other hand, the protocols based on multiple single-photon teleportation channels \cite{Andersen13,Kogias14,Marshall14} are capable of approaching the unit fidelity at finite resources, but are essentially probabilistic, because they require distinguishing all four Bell states of two photons, which is difficult at the current level of technology. The GQBS-based teleportation protocol described above relies on linear optics and photodetectors distinguishing the absence and the presence of photons (with a close to unit quantum efficiency). This protocol is deterministic for $|\alpha_0|\ge N$ and provides the unit fidelity for the circular states of one optical mode, defined by Eq.~(\ref{psi}). It means that for the states of this important class, perspective for quantum information processing \cite{Mirrahimi14,Kim15,Li17}, our teleportation scheme is preferable to any other scheme, proposed up to date. An interesting question arises: What happens if the described scheme is applied to an arbitrary state, non necessarily circular? Obviously, since the output state is always a circular one, it cannot coincide with the input state, thus the fidelity is never unit for a non-circular input. However, for an input state with the average photon number $\langle n\rangle$, choosing $N\gg\langle n\rangle$, as follows from Eq.~(\ref{cq}), one can approximate the input state by a superposition of RICS with arbitrary precision:
\begin{equation}\label{psi2}
\sum_{n=0}^{\infty}r_n|n\rangle \approx \sum_{q=0}^{N-1}r_q|\mathrm{c}_q\rangle,
\end{equation}
where $r_n$ are any coefficients. The state in the right hand side of Eq.~(\ref{psi2}) is circular and the fidelity of its teleportation can be made arbitrarily close to unity. Thus, the described teleportation protocol has high potential for quantum information processing, since it can provide an almost unit probability and almost unit fidelity at finite resources for teleportation of optical states with a sufficiently low average photon number.

\section{Pseudo-phase state and generalized Bell states}

In the regime where all coherent states in the superposition can be considered mutually orthogonal the set of these states is a discrete Fourier transform of the set of RICS and vice versa. That is why RICS can be considered as pseudo-number states and the coherent states as pseudo-phase states \cite{Kim15}. Measuring the number of photons in a RICS $|\mathrm{c}_q\rangle$ one obtains $q+lN$, as shown by Eq.~(\ref{cq}), which justifies its name of pseudo-number state. A coherent state has rather well-defined phase, which justifies its name of pseudo-phase state. Both state sets are orthonormal in this regime. At lower $|\alpha_0|$ the set of RICS remains orthonormal, but the set of coherent states $|\alpha_k\rangle$ does not. Following the approach of Pegg and Barnett \cite{Pegg89}, we can introduce a set of pseudo-phase states as the discrete Fourier transform of the set of the pseudo-number states:
\begin{equation}\label{pseudo-phase}
|\varphi_k\rangle =  \frac1{\sqrt{N}}\sum_{q=0}^{N-1}e^{-i2\pi qk/N} |\mathrm{c}_q\rangle
=\sum_{m=0}^{N-1}\tilde\delta_{mk}|\alpha_0 e^{-i2\pi m/N}\rangle,
\end{equation}
where
\begin{equation}\label{delta}
\tilde\delta_{mk} =  \frac1{N^{3/2}} \sum_{q=0}^{N-1}\frac1{\sqrt{\tilde g(q)}}e^{i2\pi (m-k)q/N},
\end{equation}
is an almost diagonal matrix, close to the Kroneker delta $\delta_{mk}$, with which it coincides at high $|\alpha_0|$, where $\tilde g(q)\to1/N$. The states $|\varphi_k\rangle$ are mutually orthogonal $\langle\varphi_k|\varphi_l\rangle=\delta_{kl}$ and create a mutually unbiased basis with RICS: $|\langle\varphi_k|\mathrm{c}_q\rangle|=1/\sqrt{N}$. Using these states as basis, we can construct the true generalized Bell states \cite{Cerf00,Horoshko07} for two optical modes
\begin{eqnarray}\label{GBS}
|\tilde\Phi_{qp}\rangle_{AB} &=& \frac1{\sqrt{N}}\sum_{m=0}^{N-1}e^{i2\pi qm/N} |\varphi_m\rangle_A|\varphi_{m+p}\rangle_B \\\nonumber
&=& \frac1{\sqrt{N}}\sum_{k=0}^{N-1} e^{-i2\pi(q-k)p/N}|\mathrm{c}_k\rangle_A|\mathrm{c}_{q-k}\rangle_B,
\end{eqnarray}
which can be used for standard deterministic teleportation of qudits. However, the way of generation of the state Eq.~(\ref{GBS}) is not clear, while a GQBS can be produced by beam-splitting a multicomponent cat state, as shown by Eq.~(\ref{out0}).

\section{Conclusion}
We have introduced a class of entangled states of two optical modes, GQBS, which are direct generalizations of Bell states of two qubits to the case of $N$-level systems encoded into superpositions of coherent states on the circle. We have shown that these states can be written in orthogonal bases where they have non-uniform coefficients. We have shown that an exact probabilistic teleportation of circular states of optical mode (optical qudits) is possible on the basis of these states and that the GQBS $|\Phi_{00}\rangle_{AB}$ is the optimal resource for this purpose. The probability of teleportation success is reasonably high for currently available sizes of cat states in the optical and the microwave domains, which allows us to hope for a possible experimental realisation of the proposed teleportation protocol. We have shown that for sufficiently high coherent amplitude and number of components this teleportation protocol can approach unit fidelity and unit probability at finite resources for any input state. We have also discussed the true generalized Bell states of two optical modes and shown that they are related to nonclassical pseudo-phase states. The obtained results can be useful for various schemes of quantum metrology and quantum information processing on the basis of information encoding in superpositions of coherent states of light.

\section*{Acknowledgments}

The authors are grateful to Stephan de Bi\`evre for many fruitful discussions. This work was supported by the European Union's Horizon 2020 research and innovation programme under grant agreement No 665148 (QCUMbER).




  \bibliography{Quasibell-noURL}
  \bibliographystyle{elsarticle-num-names}





\end{document}